\documentclass[twocolumn]{aastex631} 

\usepackage[]{mathtools}
\usepackage{bm}
\usepackage{graphicx}
\usepackage{subfigure} 
\usepackage{float}
\usepackage{verbatim}
\usepackage{booktabs}
\usepackage{multirow}
\usepackage{color}
\usepackage{mathtools}
\usepackage{longtable}
\usepackage{comment}
\usepackage{amssymb}
\usepackage{bm}

\newcommand{\Msun}{\ensuremath{\mathrm{M}_\odot}}
\begin{document}

\title{Decoding Long-duration Gravitational Waves from Binary Neutron Stars with Machine Learning: Parameter Estimation and Equations of State}
\author[0000-0002-3033-6491]{Qian Hu}
\email{Qian.Hu@glasgow.ac.uk}
\affiliation{Institute for Gravitational Research, School of Physics and Astronomy, University of Glasgow, Glasgow, G12 8QQ, United Kingdom}

\author[0000-0002-2364-2191]{Jessica Irwin}
\affiliation{Institute for Gravitational Research, School of Physics and Astronomy, University of Glasgow, Glasgow, G12 8QQ, United Kingdom}

\author[0009-0004-5204-765X]{Qi Sun}
\affiliation{Department of Computer Science, City University of Hong Kong, Tat Chee Avenue, Kowloon, Hong Kong SAR }

\author[0000-0001-7488-5022]{Christopher Messenger}
\affiliation{Institute for Gravitational Research, School of Physics and Astronomy, University of Glasgow, Glasgow, G12 8QQ, United Kingdom}

\author[0000-0003-3783-7448]{Lami Suleiman}
\affiliation{Nicholas and Lee Begovich Center for Gravitational Wave Physics and Astronomy, California State University Fullerton, Fullerton, California 92831, USA}
\def\ls{\textcolor{green}}

\author[0000-0002-1977-0019]{Ik Siong Heng}
\affiliation{Institute for Gravitational Research, School of Physics and Astronomy, University of Glasgow, Glasgow, G12 8QQ, United Kingdom}

\author[0000-0002-6508-0713]{John Veitch}
\email{John.Veitch@glasgow.ac.uk}
\affiliation{Institute for Gravitational Research, School of Physics and Astronomy, University of Glasgow, Glasgow, G12 8QQ, United Kingdom}

\date{\today}

\begin{abstract}
    Gravitational waves (GWs) from binary neutron stars (BNSs) offer valuable understanding of the nature of compact objects and hadronic matter, {and the science potential will be greatly enhanced by the third-generation (3G) GW detectors, which are expected to detect BNS signals with order-of-magnitude improvements in duration, detection rates, and signal strength. However, the resulting computational demands for analyzing such prolonged signals pose a critical challenge that existing Bayesian methods cannot feasibly address in the 3G era.
    To bridge this critical gap, }we demonstrate a machine learning-based workflow capable of producing source parameter estimation and constraints on equations of state (EOSs) for hours-long BNS signals in seconds with minimal hardware costs. We employ efficient compression of the GW data and EOS using neural networks, based on which we build normalizing flows for inference {that can deliver results in seconds. 
    The optimized computational cost of BNS signal analysis with our framework shows that machine learning has the potential to be an indispensable tool for future catalog-level BNS analyses, paving the way for large-scale investigations of BNS-related physics across the 3G observational landscape.}
\end{abstract}

\section{\label{sec0}Introduction}
The detection of gravitational waves (GWs) from binary neutron stars (BNSs)~\citep{abbott2017_GW170817ObservationGravitational, abbott2019_PropertiesBinaryNeutron, abbott2017_MultimessengerObservationsBinary} has brought valuable insights into numerous problems in fundamental physics and astrophysics~\citep{abbott2019_PropertiesBinaryNeutron,De:2018uhw,Mooley:2018qfh,Capano:2019eae,Nicholl:2017ahq, LIGOScientific:2018cki, Annala:2017llu, Kasen:2017sxr,Margalit:2017dij, LIGOScientific:2017adf, Baym:2017whm, Kasliwal:2017ngb, Annala:2019puf,Bauswein:2018bma,Dietrich:2020efo}. In particular, neutron stars (NSs) experience deformation due to the strong tidal forces during the late stages of binary {inspiral}, revealing properties of hadronic matter in their extremely dense cores which have not been probed by any other experiments or observations. This makes BNS systems ideal probes of the equation of state (EOS) of NSs, shedding light on strong nuclear interactions in extreme conditions~\citep{LIGOScientific:2018cki, Annala:2017llu, Capano:2019eae, De:2018uhw, Baym:2017whm, Annala:2019puf, Bauswein:2018bma, Dietrich:2020efo}. Proposed third-generation (3G) GW detectors, including Einstein Telescope~\citep{Punturo:2010zz, Abac:2025saz} and Cosmic Explorer~\citep{reitze2019_CosmicExplorerContribution}, are expected to detect over $2\times10^5$ BNS events per year with enhanced signal-to-noise ratios (SNRs)~\citep{Borhanian:2022czq,Branchesi:2023mws}, offering remarkable potential for groundbreaking discoveries in fundamental particle physics. 

However, a series of computationally intensive analyses are required for this purpose.
{After a BNS is detected, its} source parameters, including component masses $m_{1,2}$ and tidal deformability parameters $\Lambda_{1,2}$, must be estimated from GW data~\citep{De:2018uhw, Dietrich:2020efo}. For current detectors, this is achieved through stochastic sampling under the Bayesian inference framework~\citep{Veitch:2014wba}, which is the main bottleneck in data analysis due to its high hardware and time cost. {A typical short-duration ($<32$s), low-SNR ($<50$) signal in current detectors usually take {tens to hundreds of CPU hours to analyze using the Bilby inference package~\citep{KAGRA:2023pio,Ashton:2018jfp}}, and the time cost of parameter estimation (PE) scales up with both SNR and signal duration~\citep{Hu:2024mvn}. }
BNS signals in the 3G detectors may persist for hours due to the improved low-frequency sensitivity, rendering PE exceptionally slow. Although a number of acceleration methods have been proposed, including reducing the size of the data~\citep{Vinciguerra:2017ngf, Morisaki:2021ngj}, speeding up the likelihood evaluation~\citep{Canizares:2013ywa, Canizares:2014fya, Smith:2016qas, Cornish:2010kf, Zackay:2018qdy, Leslie:2021ssu, Smith:2013zya, Pankow:2015cra, Pathak:2022iar}, and developing more efficient samplers~\citep{Williams:2021qyt, Wong:flowmc, ElGammal:2025dkz}, limited progress has been made toward fast PE with full physics for such signals {in 3G detectors}. \citet{Smith:2021bqc} demonstrated the feasibility of using reduced-order-quadrature~\citep{Canizares:2013ywa, Canizares:2014fya, Smith:2016qas} to analyze long BNS signals and achieved an inference time of 1600 CPU hours. 
{Other works are mostly based on current detectors:} \citet{Wong:2023lgb} employed relative binning (also known as heterodyned likelihood)~\citep{Cornish:2010kf, Zackay:2018qdy, Leslie:2021ssu} and a gradient-based sampler, achieving a 3 minute inference time with GPU acceleration for aligned-spin BNS events without tidal effects using current detectors, compared with 9600 CPU hours using a traditional sampling method. This was later extended by \citet{Wouters:2024oxj} to include tidal effects, which led to a 30 minute analysis time using GPU. {\citet{Nitz:2024nhj} proposed an importance sampling-based method and predicted a BNS analysis time of $\mathcal{O}(10)$ seconds using $\mathcal{O}(10)$ cores with future optimization. }

The subsequent inference of the EOS, which depends on the source PE, also requires Bayesian stochastic sampling~\citep{LIGOScientific:2018cki}. The EOS describes the relationship between pressure ($P$) and density ($\rho$) of NS matter, which can be parameterized by piecewise polytropic~\citep{Read:2008iy} or spectral~\citep{Lindblom:2010bb} representations. Posterior samples from PE can be used to infer parameters of these EOS representations, leading to the constraints of the $P-\rho$ relation and reflecting the fundamental properties of hadronic interactions. The EOS inference alone can take tens to hundreds of CPU hours per event, bringing a further burden for 3G analysis pipelines.

{A detailed estimation of computational costs {of nested sampling} for 3G detectors~\citep{Hu:2024mvn} shows that analyzing BNS events detected in a year once could take $\mathcal{O}(10-100)$ million CPU hours per year with current acceleration methods, leading to $\mathcal{O} (10)$ GWh of electricity and millions of USD in electricity charges. }This is a substantial burden on current international computing clusters, considering that the LIGO-Virgo-KAGRA (LVK) Computing Infrastructure~\citep{bagnasco2023} has fewer than 50k CPU cores - not to mention the budget constraints and environmental impacts.

Machine learning methods have shown considerable potential in efficient GW data analysis, including signal detection~\citep{Gebhard:2019ldz, Schafer:2022dxv, Skliris:2020qax}, parameter estimation~\citep{Gabbard:2019rde, Dax:2021tsq, Langendorff:2022fzq} and subsequent
{astrophysical} analyses~\citep{McGinn:2024nkd, Stachurski:2023ntw, Ruhe:2022ddi,Colloms:2025hib}. Specifically, conditional normalizing flows (CNFs)~\citep{kobyzev2020normalizing, papamakarios2021normalizing} are widely used as density estimators to approximate the true posterior distribution, which is traditionally obtained by Bayesian sampling. A CNF is a type of neural network that learns  {conditional} differentiable and invertible transformations 
to convert the {target posterior} distribution to a {simpler latent} distribution {(e.g. a multivariate Gaussian), dependent on observed data}. 
{It can take observational data as a condition, and use it as part of the input to predict a conditional mapping between the latent and target distributions.}
During inference, samples can be rapidly drawn from the latent distribution and mapped back to the target posterior given 
{a} condition. A series of its successful applications in GW parameter estimation and subsequent probes into fundamental physics, primarily focusing on current detectors or short duration signals, can be found in \citet{Green:2020dnx, Dax:2021tsq, Dax:2022pxd, Gupte:2024jfe, Dax:2024mcn}. Notably, \citet{Dax:2024mcn}, applies CNFs to BNS signals, which is made possible using heterodyning and multibanding, along with a prior conditioning algorithm to train networks adaptable to different mass priors. We adopt a similar framework, but focus on longer signals and higher SNRs which represent the most challenging case in 3G detectors.

\begin{figure}
  \includegraphics[width=0.48\textwidth]{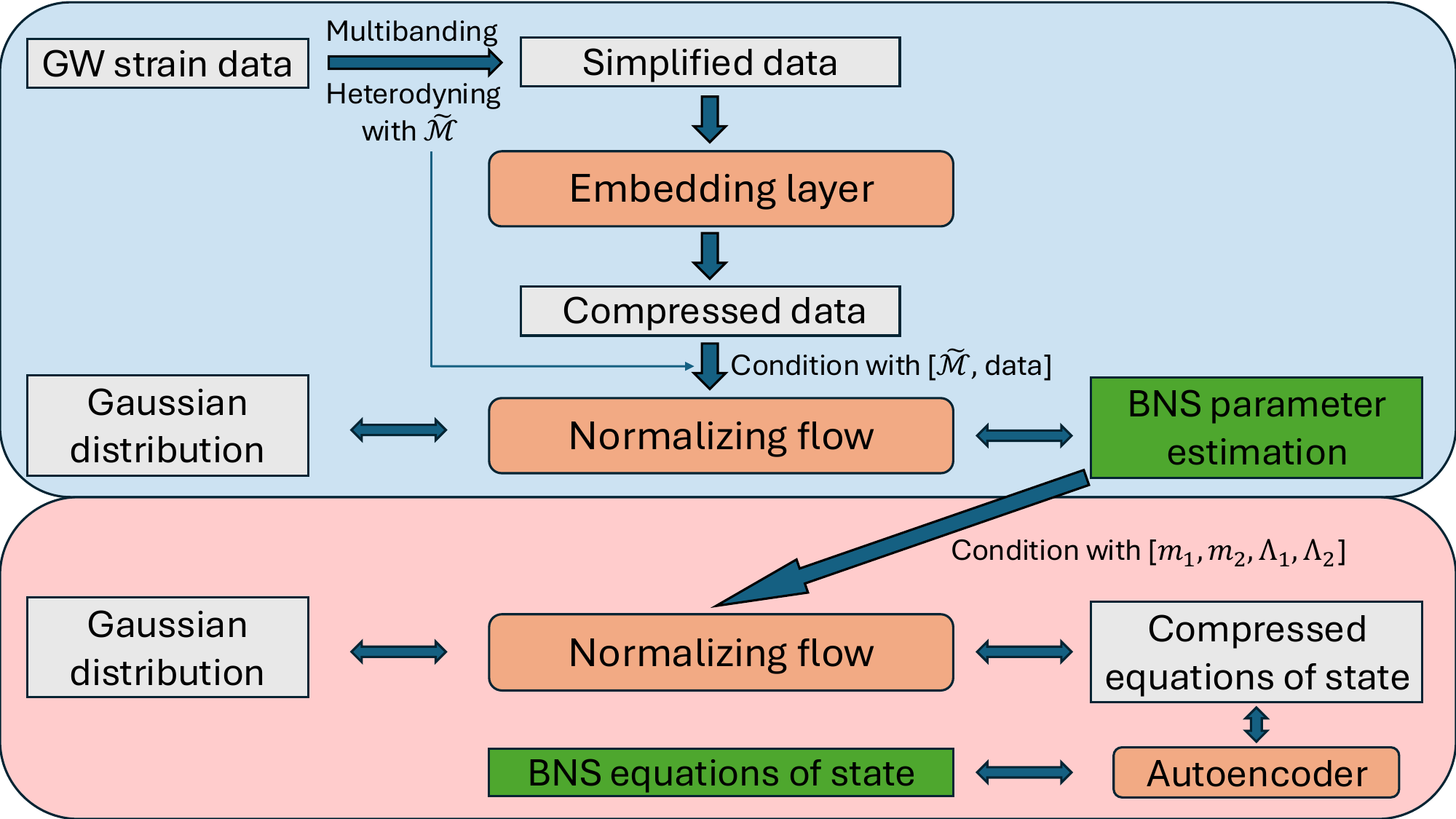}
  \centering
  \caption{\label{fig:net_struc} Structure of our flow-based BNS analysis pipeline. The normalizing flows learn transforms between a Gaussian distribution and the target posterior distribution. The first flow in conditioned on the preprocessed (via multibanding and heterodyning) and compressed (by a neural network) GW strain data and generate source parameter estimation, which then conditions the second flow to generate a reduced representation of NS EOS. The reduced representation and original EOS can be transformed into each other by an neural auto-encoder. }
\end{figure}

In this work, we develop a CNF-based analysis pipeline capable of generating full parameter estimation (with GW phase marginalized) for long signals from precessing BNSs in the 3G GW detectors, and generating EOS estimation within a second on a single GPU. The structure of the pipeline is shown in Fig.~\ref{fig:net_struc}. The first CNF is conditioned on the preprocessed and compressed GW strain data to generate source PE, and the second CNF uses the PE results to infer the NS EOS. Details are provided below. 

\section{Building inference models}
\subsection{\label{sec1}Parameter space (prior)}
BNS parameters, especially the chirp mass, can be tightly constrained by 3G detectors~\citep{Borhanian:2022czq,Branchesi:2023mws}. We find it is extremely challenging to train a model to cover a wide range of chirp mass and SNR, as the posterior is much narrower than the prior (training parameter space). To address this, we divide the parameter space to several regions and train a parameter estimation model for each. As an example, we consider two SNR ranges, 20-50 (low SNR) and 200-500 (high SNR). The former includes the majority of relatively informative BNS events in the 3G era, while the latter represents the louder ones. The detector frame chirp mass is sampled uniformly between $2\,\Msun$ and $2.1\,\Msun$ for low SNR prior, and $1.3\,\Msun$ and $1.31\,\Msun$ for high SNR prior. Given our detector configuration - one triangular ET at ET-D sensitivity~\citep{Hild:2010id} at {the existing} Virgo site and two 40-km CEs at {CE2} design sensitivity~\citep{reitze2019_CosmicExplorerContribution} at Hanford and Livingston~\citep{LIGOScientific:2016wof} - these mass and SNR ranges correspond to source-frame chirp masses consistent with current knowledge of BNS systems. {The redshifts of low-SNR and high-SNR BNSs are $\sim 0.1$ and $\sim 0.8$, respectively.} The mass ratio $q=m_2/m_1$ is sampled uniformly between 0.5 and 1. The dimensionless spins are isotropic with the maximum magnitude of 0.05. Tidal {deformability} parameters are parameterized as $\tilde{\Lambda}$ and $\delta \tilde{\Lambda}$~\citep{PhysRevD.89.103012}, where $\tilde{\Lambda}$ is uniform between 0 and 1600 {following the estimates from GW170817~\citep{abbott2019_PropertiesBinaryNeutron}}, and $\delta \tilde{\Lambda}$ is uniform within its allowed range, which is determined by component masses and $\tilde{\Lambda}$ such that both components have positive deformability parameters. {A wider prior for tidal parameters would benefit the applicability of our model in low mass scenarios, but here we start with 1600 as a proof-of-concept.} The prior for extrinsic parameters follows common choices except for luminosity distance $d_L$: We sample the SNR uniformly in the target SNR band, then scale the $d_L$ accordingly. This ensures the models consistently encounter similar loudness levels, reducing the number of outliers that are difficult to learn. Although our prior distribution is not perfectly aligned with those used in traditional parameter estimation, this data-independent discrepancy can be removed by importance sampling~\citep{Dax:2022pxd, Nitz:2024nhj}.

\subsection{\label{sec2}Data preprocessing and compression}
\begin{figure}
  \includegraphics[width=0.48\textwidth]{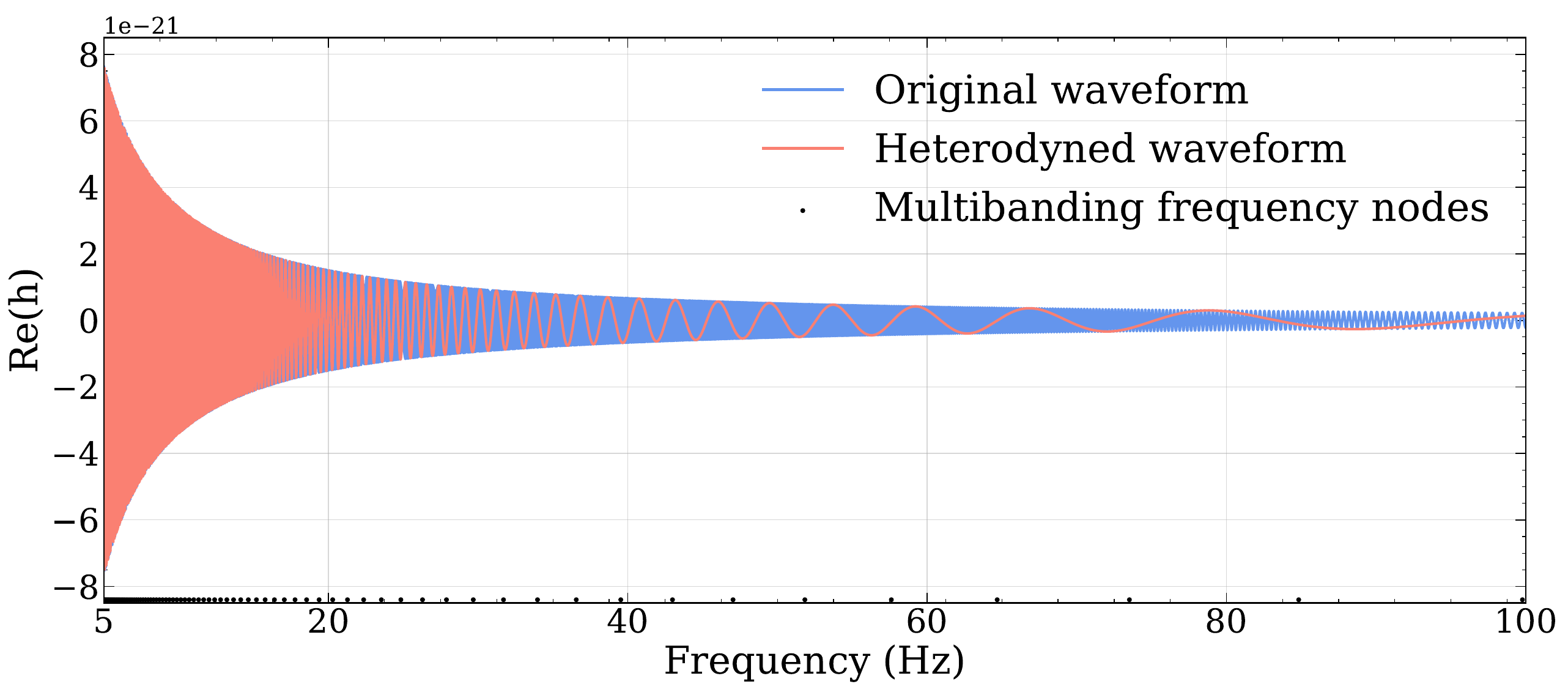}
  \caption{\label{fig:hd_waveform} BNS frequency-domain waveforms before (blue) and after (orange) heterodyning. Waveforms are truncated at 100\,Hz for better illustration. Black dots represent the frequency nodes of the multibanding scheme, which bands defined between nodes. }
\end{figure}

Considering a frequency band starting from 5\,Hz, a three-hour-long BNS signal ($\sim 1.1+1.1\,\Msun$) with sampling rate of 2048\,Hz has roughly 12 million data points, {far too many to be used in traditional sampling algorithms {or machine learning algorithms}.}
As the GW frequency increases during inspiral, a multibanding scheme~\citep{Vinciguerra:2017ngf, Morisaki:2021ngj}, i.e. adaptively adjusting the sampling frequency, is an effective way of reducing the data size and accelerating analyses. We propose a novel multi-banding scheme that adaptively selects frequency nodes and resolutions, ensuring that each band's resolution is precisely tuned to the needs of BNS signals.
In particular, we divide the full bandwidth into bands containing {roughly $N=64$ data points each}, and search from the high frequency cut-off $f_0$ (1024Hz for this work) {to lower frequencies until the first frequency $f_1$} such that 
\begin{equation}
    \label{eq:mbcond}
   \alpha_\mathrm{safety} (f_0-f_1)  \left[ \tau(f_1) - \tau(f_0) \right] > N,
\end{equation}
where $\tau(f)$ is the time-to-merger function to 3.5 post-Newtonian order~\citep{Buonanno:2009zt}, and $\alpha_\mathrm{safety}=2$ is a safety factor that enlarges the effective band duration, ensuring that the frequency resolution is high enough to cope with different source parameters and the potential errors in $\tau(f)$. The frequency resolution in the band $(f_1, f_0]$ is given by
\begin{equation}
    \label{eq:mbdf}
    \Delta f_0 = \frac{1}{\alpha_\mathrm{safety} \left[ \tau(f_1) - \tau(f_0) \right]},
\end{equation}
which corresponds to $T_0=\alpha_\mathrm{safety} \left[ \tau(f_1) - \tau(f_0) \right]$ data in the time domain. 
This process is repeated to obtain $f_2,~f_3,~\dots$ and the corresponding $\Delta f_i$, until the lower frequency bound reaches the 5\,Hz cut-off. This scheme reduces the 12 million data points to approximately 6000 {in 81 bands}. The selected frequency nodes are shown as black dots in Fig.~\ref{fig:hd_waveform}.

Singular Value Decomposition (SVD) is used to further compress the data by projecting it onto linear bases that represent the waveform space. However, a BNS waveform is highly oscillatory in the frequency domain, making SVD inefficient. Following \citet{Dax:2024mcn}, we employ heterodyning (relative binning)~\citep{Cornish:2010kf, Zackay:2018qdy, Leslie:2021ssu} to reduce the oscillation in the data (Fig.~\ref{fig:hd_waveform}). Specifically, we multiply the signal by a phase factor $e^{ i \frac{3}{128 } (\frac{\pi G\mathcal{M} f}{c^3})^{-5/3}}$, which is the inverse of the leading term of the oscillations in the GW waveform. This effectively smooths the waveform and improves the compression ratio of the SVD. With our restricted prior, we can retain the first 128 SVD bases sorted by singular values and reconstruct the zero-noise signal with a median mismatch of $\sim 10^{-8}$ and maximum mismatch of $\sim 10^{-6}$. 

Neural networks are then used to further compress the SVD projection of data from multiple detectors. For cross-validation, we explore two types of neural networks for embedding: a residual network~\citep{he2015deep} of Multi-Layer Perceptrons (MLPResNet) and a Vision Transformer (ViT)~\citep{dosovitskiy2021imageworth16x16words}. However, due to the larger memory requirement of ViT, we only perform the cross-validation in low-SNR models. The five data streams from the 1ET+2CE network are compressed into a vector of length of 128, which is used to condition CNF for parameter estimation. {We note that the compression by the embedding layer is not as directly verifiable as SVD. Instead, it is trained together with the CNF and assessed by the final PE results. }

\subsection{Training the parameter estimation model}
The training set should comprehensively cover the parameter space, which, in analogy to template bank generation~\citep{PhysRevD.44.3819, Brown:2012qf, Harry:2016ijz, Nitz:2021vqh}, becomes more challenging in low mass, high SNR and high dimension scenarios. Meeting all these factors, we find that large amount of training data is necessary to avoid overfitting. 

GW strain data (signal and random Gaussian noise) is simulated on-the-fly during training using the waveform model \texttt{IMRPhenomPv2\_NRTidal}~\citep{Dietrich:2019kaq}. Sky location parameters, coalescence time and SNR are randomly generated for each simulation, ensuring an infinite number of possible samples. The remaining 12 parameters are drawn from the prior before training and are used to calculate GW waveform.  We save the SVD projections of the waveforms for loading during training. A total of 64 million {waveforms} are used to train low-SNR models, and 100 million {waveforms} are used for high-SNR models. The CNF model used for PE is a neural spline flow~\citep{NEURIPS2019_7ac71d43}. {Each PE model (the embedding network after SVD and the CNF) has roughly 160 million learnable parameters with 96 million belonging to the normalizing flow, and all parameters optimized jointly by minimizing the negative log-likelihood.  We train the networks for 2-3 weeks on an NVIDIA A100 GPU. Further technical details and discussions are given in \citet{QianThesis}. }

{The frequency-dependent response to GWs is important to the accuracy and precision of data analysis in 3G detectors~\citep{Zhao:2017cbb, Hu:2023hos, Rakhmanov:2008is, Essick:2017wyl}, but it introduces significant computational challenges~\citep{Baker:2025taj}. We demonstrate that machine learning algorithms have the potential to address these challenges. Focusing on long signals, we include the effect of Earth's rotation that modulates the low-frequency response when generating the training data. Under the stationary phase approximation, the frequency-domain detector response is written as 
{
\begin{equation}
    \begin{aligned}
    h(f) = \sum_{A=\{+,\times\}}F_{A}\left[t(f)\right] h_A(f)e^{-2\pi i f [t(f)+\Delta t(f)]} ,
    \end{aligned}
\end{equation}
where $F_{A}$ is the antenna response function and $h_A(f)$ is the waveform polarization in the frequency domain. $t_c$ is the coalescence time, $t(f) = t_c-\tau(f)$, and $\tau(f)$ is the same as used in Eq.~\ref{eq:mbcond}. Since $t_c$ and $t(f)$ are measured at geocenter, an additional factor $\Delta t$ that accounts for the GW arrival time delay between the geocenter and the detector is required, which becomes frequency-dependent when the Earth rotation is included. Sky location dependency of $F_{A}$ and $\Delta t$ is omitted in the equation for brevity.} The high-frequency modulation caused by the detector's free spectral range~\citep{Rakhmanov:2008is, Baker:2025taj,Baral:2025geo}, while in principle can be included in the same way as the Earth's rotation effects, is ignored in this work as we set a high frequency cutoff at 1024\,Hz. The full frequency-dependent response is planned for future work.
}

The chirp mass used for heterodyning is assumed to be perfectly known when obtaining SVD bases, but it is not known in inference. Following \citet{Dax:2024mcn}, 
during training, the GW data is heterodyned with $\tilde{\mathcal{M}} = \mathcal{M} + \delta \mathcal{M}$ instead of the exact chirp mass $\mathcal{M}$, where small perturbation $\delta \mathcal{M}$ enables the network to deal with inaccuracies in heterodyning. We set $\delta \mathcal{M}$ uniformly distributed in $[-0.0005, 0.0005]\,\Msun$ for low-SNR models and $[-0.0001, 0.0001]\,\Msun$ for high-SNR models. $\tilde{\mathcal{M}}$ is also given as a condition to the normalizing flow. During inference, the entire chirp mass prior can be divided into several segments of length 0.001 (or 0.0002) $\Msun$, with the segment yielding the highest likelihood being selected. 

\section{Results and validation}
\subsection{Parameter estimation}
We infer 16 out of the total 17 BNS parameters, with the coalescence phase marginalized by excluding it from the normalizing flow while still incorporating its underlying variations in the data. {We then perform importance sampling based on the framework proposed in \citet{Dax:2022pxd} to refine the estimation. More specifically, each sample $\theta$ drawn from the CNF is assigned a weight $p(\theta |d)/q(\theta|d)$, where $p$ is the exact posterior function in PE and $q$ is the approximate probability density function from the CNF. For validation, we run full Bayesian PE with the Earth rotation included using our multibanding scheme and the \texttt{dynesty} sampler~\citep{Speagle_2020} implemented in \texttt{bilby}~\citep{Ashton:2018jfp} as the ground truth distribution. As an example, we simulate a BNS event of network SNR 35 and analyze it with our low-SNR models and \texttt{bilby}. Our model can accomplish sampling within a second on an NVIDIA 3080Ti GPU and the importance sampling costs $\mathcal{O}(1)-\mathcal{O}(10)$ CPU hours, depending on the sampling efficiency. By contrast, the full PE requires 3500 CPU hours. }

\begin{figure*}
  \includegraphics[width=0.98\textwidth]{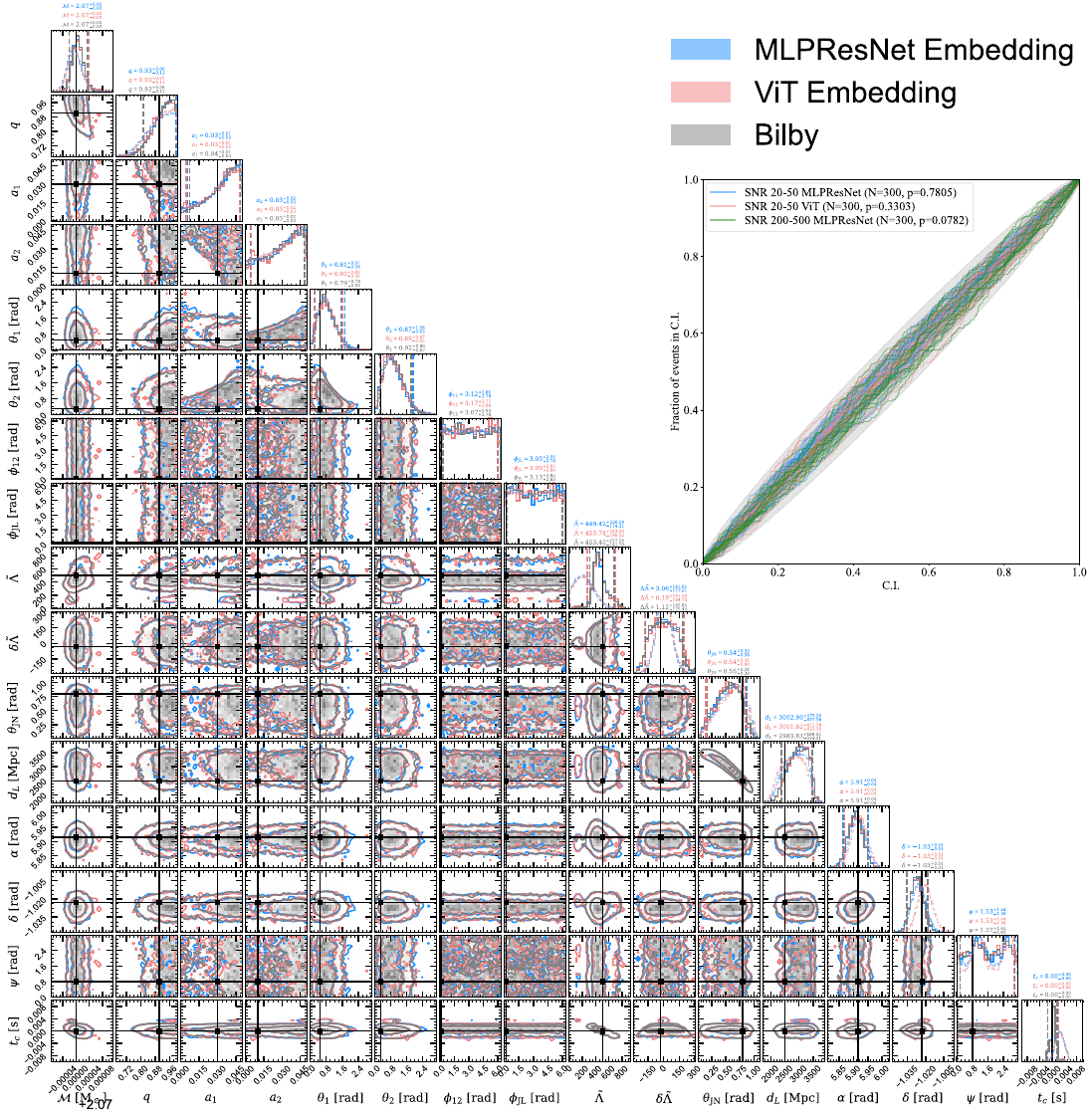}
  \caption{\label{fig:example_corner} {Lower left: Corner plots for an example BNS event. The blue region shows the posterior distribution generated by MLPResNet embedded model and the red is from ViT model after importance sampling. The initial samples from the two models are shown in dashed lines in the 1D marginal distributions and they are not shown in the 2D contours. Grey represents the ground truth result obtained by \texttt{bilby}. The source is a $1.74+1.56\,\Msun$ system located at 2500\,Mpc with a network SNR of 35. The true values of the source parameters are shown as black lines, and dashed vertical lines in 1D distributions show the 5\% and 95\% {credible intervals}. The chirp mass is presented in the detector frame. Middle right: P-P plots of all PE models. $x$-axis is the credible interval and $y$-axis the fraction of events included in the corresponding credible interval. Grey confidence regions are of 68\%, 95\%, 99.7\% confidence levels. Different models are represented by different colors, while parameters for each model are plotted with the same color. $p$-values of different parameters are combined and are shown in the legend{, and indicate no significant statistical biases in the results.}
  }}
\end{figure*}

{The initial samples from our models, weighted samples after importance sampling, and \texttt{bilby} samples are shown in Fig.~\ref{fig:example_corner}. While the initial samples agree across our models, they show differences with the ground truth, especially in the tidal deformability parameters. This behavior is expected, because the normalizing flows are approximations to the true probability density, and we have used a considerably different prior in tidal parameters during training for faster training convergence. We note that, among many simulations we have performed, the difference is not always as distinct as shown here - Fig.~\ref{fig:example_corner} actually presents a bad case. The differences are mostly removed after importance sampling, and the weighted samples show negligible differences with the \texttt{bilby} result. Therefore, our models are able to provide accurate posteriors at a small computational cost.}


The self-consistency of our PE models is assessed by the p-p test~\citep{PhysRevD.89.084060}, in the sense that $x\%$ confidence interval should successfully predict $x\%$ of events. {Without performing importance sampling,} the p-p plots are shown in {the right side of} Fig.~\ref{fig:example_corner}. All models pass the p-p test. {As importance sampling guarantees to improve posterior accuracy, we expect the self-consistency to hold after importance sampling. }


\subsection{Constraining equations of state}
Following \citet{McGinn:2024nkd}, we train another CNF based on RealNVP~\citep{dinh2017densityestimationusingreal} to infer the NS EOS. 
{The EOS used for training is generated using the publicly available code CUTER~\citep{Davis:2024nda}, which uses a semi-agnostic approach with a meta-model constrained by nuclear theory at low density, and piecewise polytropes at high density.}
We generate physically plausible BNS source parameters $\{m_1, m_2, \Lambda_1, \Lambda_2 \}$ from each EOS. The CNF, conditioned on the BNS source parameters, is trained to infer a 12 dimensional representation of the EOS compressed by a convolutional auto-encoder, along with hyperparameters of each EOS, including {maximum allowed pressure and energy density}. The {trained} auto-encoder can then reconstruct the $P-\rho$ EOS {by} providing the 12 dimensional representation. 

\begin{figure}
  \includegraphics[width=0.48\textwidth]{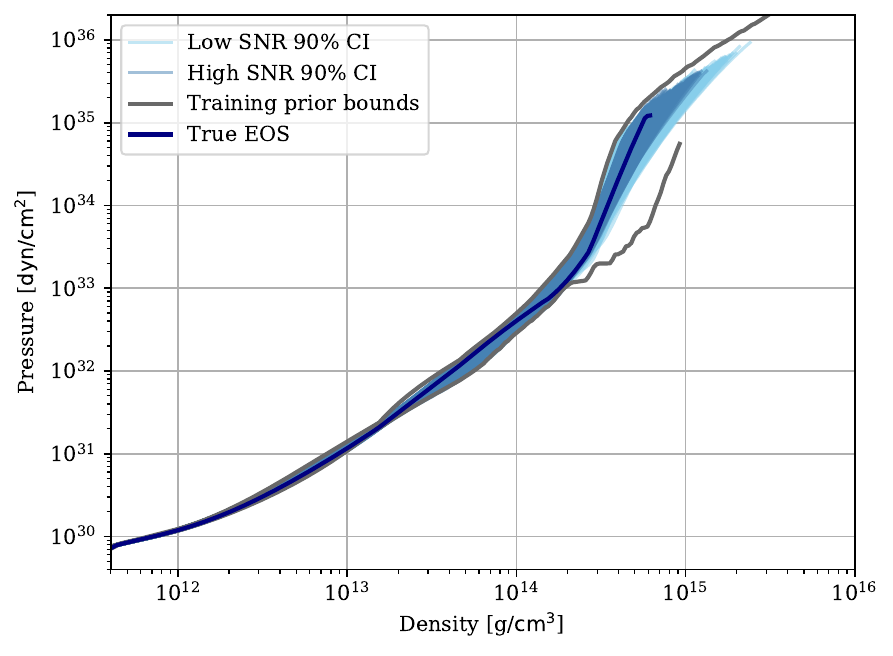}
  \caption{\label{fig:eos} 90\% confidence intervals (CIs) of EOS constraints for simulated BNS events. The CIs are defined in the 12D representation space and mapped to pressure-density plane by the auto-encoder. Two BNS events, governed by the same EOS, have different network SNRs: 39 (low-SNR, in light blue) and 390 (high-SNR, in blue). The injected EOS is given in dark blue with the training prior bounds given in grey, illustrating the most stiff and most soft EOSs in our training data set. }
\end{figure}

We simulate two BNS events with identical underlying EOS but differing network SNRs of 39 and 390 and then perform parameter estimation with our PE models. The resulting posterior samples {(converted to source frame using a $\Lambda$CDM cosmology model informed by the Planck observatory data of 2015~\citep{Planck:2015fie})} are passed to the EOS inference model to generate EOS posteriors (samples of 12D representation plus hyperparameters), which are then decoded into pressure-density relationships using the auto-encoder (Fig.~\ref{fig:eos}). We successfully recover the underlying EOS in both cases, with tighter constraints in the high-SNR scenario as expected. The EOS inference takes less than one second, greatly {reducing} the computation cost compared with traditional stochastic sampling, {as initially demonstrated in \citet{McGinn:2024nkd}.}

{We note that combining multiple events would significantly tighten the EOS constraints, as the true underlying EOS would pass posterior parameter regions of all BNS events. Intuitively, this is a more effective way to improve EOS constraints than improving the SNR, because a single tight posterior region still allows a certain level of flexibility in EOS. We also note that the EOS could be constrained better with a higher frequency cutoff in PE. We leave these improvements to our future work.}

\section{Discussions}
We demonstrate a reliable machine learning based analysis pipeline for long BNS signals in the 3G GW detectors, {showing that the advantages of current machine learning methods not only still hold, but are also amplified for the toughest problems in the future}. While certain tasks for long BNS signals could be prohibitively slow for traditional methods, our approach processes each event in under a second on a single GPU. {Assuming $\mathcal{O}(100)$~W power for a GPU, our method would only use $\mathcal{O}(100)$kWh electricity in inference and $\mathcal{O}(10)$~MWh in training, reducing the total cost of data analysis by a factor of 1/1000. The factor changes to 1/100 when importance sampling is used as a further refinement, but it is still an crucial improvement that enables catalog-level analyses in the 3G era.}
{As a showcase, we have presented a feasible and exciting opportunity to advance population-level knowledge of hadronic matter and NS, which is also accelerated by machine learning.
Outputs of other follow-up studies of topics such as cosmology~\citep{LIGOScientific:2017adf, DES:2019ccw, LIGOScientific:2019zcs}, the astrophysical population~\citep{LIGOScientific:2018jsj, LIGOScientific:2020kqk, KAGRA:2021duu}, and the stochastic GW background~\citep{LIGOScientific:2016jlg, LIGOScientific:2016nwa, LIGOScientific:2017zlf} will be greatly enhanced by the million-events-level catalogs in the 3G era, and a framework to rapidly generate PE samples and translate PE to physical implications is necessary to ensure the feasibility of these studies.}



Looking ahead, there is also a pressing need to adapt these algorithms to more realistic scenarios expected in the 3G detectors. This includes addressing challenges including time variations in noise and overlapping signals, {improving the efficiency of training}, etc~\citep{himemoto2021_ImpactsOverlappingGravitationalwave,pizzati2022_InferenceOverlappingGravitational,relton2021_ParameterEstimationBias,relton2022_AddressingChallengesDetecting,samajdar2021_BiasesParameterEstimation,Hu:2022bji}, ensuring that the analysis methods remain robust and reliable under all conditions and {easy-accessible to the community}. {These improvements and challenges will be explored in our future work.}

\begin{acknowledgments}
  The authors would like to thank Nihar Gupte, Maximilian Dax, Xue-Ting Zhang, and Michael Puerrer for helpful discussions and suggestions. The authors are grateful for computational resources provided by the LIGO Lab at Caltech which is supported by National Science Foundation Grants PHY-0757058 and PHY-0823459. QH is supported by STFC grant ST/Y004256/1 and CSC. JV is supported by STFC grant ST/Y004256/1.

\end{acknowledgments}


\bibliography{refs.bib}


\end{document}